# A Framework to Prevent QR Code Based Phishing Attacks

T.T. Dayaratne
WSO2
Sri Lanka
thusithathilina@gmail.com

*Abstract*—**Though the rapid development and spread of Information and Communication Technology (ICT) making people's life much more easier, on the other hand it causing some serious threats to the society. Phishing is one of the most common cyber threat, that most users falls in. This research investigate on QR code based phishing attacks which is a newly adopted intrusive method and how to enhance the awareness and avoidance behavior of QR based phishing attacks through the user centric security education approaches using game based learning.**

*Keywords-component; phishing; QR codes; learning.*

## I. INTRODUCTION

The penetration of the Internet and smartphones has enable ample of opportunities, where its' users can shop, communicate, do payments, etc... with few clicks or taps[1]. That makes smartphones and the internet are essential parts of everyday life. With more and more people using internet and smartphones, intruders are trying to target these audience with malicious intents.

Social engineering techniques along with phishing is one of the most recurrent cyber threat, that people can easily falls in[2]. Phishing is considered as a semantic attack and commonly interpreted as online identity theft[3]. Masquerading as a trustworthy entity in order to capture sensitive data of a particular user is not a new phenomena. Scam emails which directs users to fraudulent website is the commonly used approach for phishing. But that has come to a new dimension along with the new developments in IT world. Phishing attacks based on QR(Quick Response) codes is the most recent oxygenating factor for the attackers.

QR code was originated in Japan[4] originally to track the automotive components in the industry. But with the popularity it among various industries, make a rapid development for the original QR code and now it can hold link, plain text, SMS text message, addresses, URLs, Geolocation, email, phone numbers or contact information. High information density and robustness makes 2-dimensional barcodes known as QR Codes, a popular choice among various industries that appear in more places in the urban environment for various purposes over traditional bar codes. Marketing and online-payments[5] are 2 most common industries which uses QR codes heavily. URL encoding to make information instantly available is the most common use case among these industries. QR Codes can be described as paper-based hyperlinks, which directs users to websites. Since it provides a way to access a brand's website more quickly than by manually entering a URL, according to the e marketers and google trends [6], [7] this been adopted over millions of smartphone users. QR codes also allows marketers to target their desired groups on specific locations.

Though the properties of QR code makes marketers life much easier and effective, on other hand security vulnerabilities exist itself in code and in its' readers yields QR codes to be used as a novel tool for phishing attacks.

## II. STATE OF ART – KEY LITERATURE

The fundamental misbelief of the smartphone users, that the smartphone is safer than a typical PC is a recipe for disaster. Due to this misbelief and for convenience and ease of use, users and even developers have overlooked many of the lessons learned from the past relate to phishing [8]. Thompson and Lee in their study highlighted that users make different insecure choices when the context of the situation changes, if user have to make a decision as part of the solution for a security weakness. Further they have shown that since QR codes are often embedded in physical objects such as posters, billboards, that increase users' perception of safety as they feel that they are sort of real and tangible thing instead of an untrusted website link. This leads users are more vulnerable to QR based phishing attacks than the traditional phishing methods.

Several studies have shown that QR codes can be used as an attack vector for many security threats. Keiseberg et al have conducted a proof of concept phishing attack using QR code. Further in their study they have shown dangers of possible attacks utilizing manipulated QR codes and highlighted that proper input sanitization is need to be performed prior to processing the contained data [9]. Amin et al in their study of malicious QR codes in wild state that they have found about 150 malicious QR codes that were designed to direct victims to phishing sites or direct users to either exploit or intermediate sites. Spoofed versions of password-protected websites, fake versions of the Google Play app market, malware distributed via direct download links were the common threats [10]. Though the number is quite less the consequences and impact can't be simply ignored.

Vidas et al. carried out two experiments to understand the impact of QR based phishing which they have named as QRshing in the city of Pittsburgh. In QRishing experiment they found that curiosity is the main motivation for smartphone users to scan a code[11]. A similar experiment was also conducted by Seeburger et al[12]. Along with those 2 experiments and Adrian and Katharina in their QR Inception: Barcode-in-Barcode Attacks[13] research, highlight the need for further research on adequate tools to support the smartphone user to detect potential threats in QR codes. Research also highlighted that most QR code readers do not provide feasible tools to automatically detect malicious intents embedded in QR codes .

Keiseberg et al in their QR Code Security: A Survey of Attacks and Challenges for Usable Security research, defined specific requirements which required to develop a multi-layer guidelines as a first step toward the development of a secure QR code processing environment. They have also highlighted that the usable security design guidelines and security awareness as open research challenges in the field [14].

People have also investigate on user centric security education approaches in order to designing proper systems to prevent phishing attacks. Arachchilage & Love have conducted a research on computer user's knowledge in order to prevent phishing attacks[15]. In their study, they have shown that lack of knowledge to prevent phishing threats, cause users more susceptible for attacks and educational games can be used to educate people in order to thwart phishing attacks. In [1, 16, 17, 18] they evaluated the effectiveness of a mobile game in order to protect computer users against phishing attacks. From the results it can be clearly seen that gaming based learning approaches are much effective than traditional approaches. Also user centric approaches can be used to enhance the users' behaviour by motivating them to protect themselves from phishing attacks.

However, though a considerable amount of research are being conducted in the field of phishing attacks and user centric approaches to prevent phishing attacks, since QR based phishing attack is a quite new phenomena, a very few research have been carried out on this field. And the conducted research has also highlights that, computer users are still the weakest link when comes to information security. Therefore there is a lack of a proper study which investigate on how to use user centric security education in order to prevent the QR code based phishing attacks.

In order to fill that gap, this research will investigate on ways of enhancing the awareness and avoidance behavior of QR based phishing attacks, through user centric security education approaches[18], using game based learning.

## III. RESEARCH OBJECTIVE & METHODOLOGY

The objective of research is to identify and design a novel game design framework with quantitative and qualitative analysis that can be used to mitigate the risk of QR based phishing attacks.

The research procedure would be as follows.

Existing work and ways of QR codes can be used as attack vector for phishing attacks would be investigate as the initial step of the research, in order to get a deep understanding about the topic. With the knowledge obtain by reviewing the existing work, a qualitative analysis will be carried out. This would be a laboratory study in order to identify and understand the users perspective relates to the problem. The plan is to recruit about 50 voluntary participants for the study. They will be provided a questionnaire to identify their motivation factors to scan QR codes in public.

As the the second phase, a quantitative analysis will be carry out to identify the victims of QR code based phishing attacks. In order to do so. Set of QR codes would be placed in public and some of those QR codes would be manipulate to redirect users to phihing websites, where simulated phishing attacks will be conducted while carefully observing the users' behaviors. Users will be requested to scan QR codes without prior notice about the consequences. In this phase we are guaranteeing that no actual rather simulated attacks.

Set of guidelines for secure usage of QR code will identify and developed based on the user experiment results, identifying main concerns/usages which will be used to secure the QR code usage in terms phishing attacks is the main objective of this phase. The next phase is to develop a framework to prevent the QR based phishing attacks, based on the guidelines that are being identified in prior phase. Guidelines will be further improved and fine tuned with the help of security and usability experts in the filed and used in the framework. As the last phase, a educational game will be develop using the identified framework in order to raise awareness and avoidance behavior of QR based phishing attacks and it would be evaluated to investigate the successiveness of the developed framework. Evaluation process will be carried out as a quantitative analysis through a user study. Two controlled user groups will be used to evaluate the framework, where one group will be exposed to the identified framework while other group will not. Effectiveness of the framework will be evaluated based on the result, of this study. Finally the experiment and the results will be presented through the thesis.

The research will be carry out through the following 6 phases.

1. Review and Analysis Phase (Month 1-8)
2. Discovery Phase (Month 9-18)
3. Development phase I (Month 19-23)
4. Development phase II (Month 24-27)
5. Evaluation phase (Month 28-32)
6. Thesis write-up and publication (Month 33-36)